\documentclass[aps, prb, twocolumn,floatfix, showpacs, 10pt]{revtex4}
\usepackage{amsmath}
\usepackage{amsfonts}
\usepackage{amsbsy}
\usepackage{amssymb}
\usepackage{graphicx}
\usepackage{textcomp}
\usepackage[caption=false,singlelinecheck=false]{subfig}
\usepackage{xcolor}
\usepackage{mathrsfs}
\usepackage{bm}
\usepackage{ulem}
\usepackage[colorlinks,linkcolor=red,citecolor=blue,filecolor=magenta,
urlcolor=cyan]{hyperref}

\allowdisplaybreaks

\begin{document}
\title{Translational symmetry breaking and the disintegration of the Hofstadter butterfly}

\author{Archana Mishra}

\author{S. R. Hassan}

\author{R. Shankar}

\affiliation{The Institute of Mathematical Sciences, C.I.T. Campus, Chennai 600 
113, India}

\date{\today}
\begin{abstract}

We study the effect of interactions on the Hofstadter butterfly of the
honeycomb lattice.  We show that the interactions induce charge ordering that
breaks the translational and rotational symmetries of the system. These phase
transitions are prolific and occur at many values of the flux and particle
density. The breaking of the translational symmetry introduces a new length
scale in the problem and this affects the energy band diagram resulting in the
disintegration of the fractal structure in the energy flux plot, the Hofstadter
butterfly. This disintegration increases with increase  in the interaction
strength.  Many of these phase transitions are accompanied with change in the
Hall conductivity. Consequently, the disintegration of the Hofstadter butterfly
is manifested in the Landau fan diagram also.

\end{abstract}
\maketitle
\section{Introduction}

The two dimensional electron gas (2DEG) in the presence of magnetic field has
been of special  interest to the condensed  matter  physicists since the
discovery  of the quantum Hall  effect \cite{klitzing1980}  and the fractional
quantum Hall effect \cite{tsui1982}. The 2DEG in a periodic potential and
magnetic field has been the cradle of several interesting and important
theoretical concepts like the identification of a topological invariant with
the Hall conductivity \cite{thouless1982, kohmoto1985} and the existence of a
fractal structure in the energy gaps, the Hofstadter butterfly
\cite{hofstadter1976, wannier1978}.  Interest in this phenomenon has been
recently revived with the experimental observation of  the Hofstadter butterfly
in graphene superlattices \cite{ponomarenko2013, dean2013, hunt2013, yu2014,
yu12014} and the realization of the Hofstadter Hamiltonian in the optical
lattice systems \cite{aidelsburger2013, miyake2013}. This has motivated us to study the effect of interactions  on the Hofstadter
butterfly. 

The Hofstadter butterfly emerges from the interplay of
two length scales in the lattice systems: the periodicity of the lattice and the
magnetic length \cite{hofstadter1976,wannier1978,donald1983}. It is an effect that emerges when we look at the single
particle energy gaps and Hall conductivities of a large number of physical systems, in
principle, an infinite number of systems with flux per plaquette, $\phi=p/q$,
where $p,q$ are co-prime integers and $p<q$. For non-interacting systems with fermion
densities corresponding to filled bands, the gaps exhibit a self similar
structure. 
Further, if the Hall conductivities, in units of $e^2/h$ are plotted as a function of $\phi$ and the
number of fermions per unit cell, the
contours with the same Hall conductivities are straight lines with integer
intercepts. The Hall conductivity of each contour and the intercept are
solutions of a Diophantine equation. This plot, the Landau fan diagram, is 
the experimental evidence of the Hofstadter butterfly in graphene superlattices
\cite{ponomarenko2013, dean2013, hunt2013, yu2014,
yu12014}.

Interaction can induce charge ordering that can break the translational symmetry of the system.
Interaction induced translational symmetry breaking phases have been studied previously in
the honeycomb lattice in the absence of magnetic field \cite{weeks2010,hou2007,castro2011,castro2013}.
Effect of interactions on the Hofstadter butterfly has been studied in the past
using  mean field  approximation in both square lattice \cite{gudmundsson1995,doh1998} 
and honeycomb lattice (for Dirac fermions)\cite{apalkov2014, chakraborty2013}. Electron electron interaction
in square lattice in magnetic field has also been studied using exact diagonalization method \cite{czajka2006}. However,
none  of these past works  consider translational symmetry breaking and its
effect on the fractal structure of the Hofstadter butterfly. 
Moreover, the effect
of interactions on the experimental probe of the  Hofstadter butterfly, the Landau
fan diagram, has  not been mentioned in previous work.

In interacting systems, the translational symmetry breaking introduces a third length scale in the problem
which can affect the self similarity of the Hofstadter butterfly. In this paper, we investigate whether this actually happens. 
In  our recent  paper \cite{archana2016}, we had studied the effect of interaction
in one of the many systems required to realize the Hofstadter butterfly in honeycomb lattice, lattice with
flux per plaquette $\phi=1/3$ (in units of $h/e$).
We studied the interaction induced  translational symmetry broken phases  in the  Hofstadter
regime  of the honeycomb lattice for this flux value.
In this paper, we study the effect of interactions on the spinless fermions of the honeycomb
lattice in the presence of magnetic field such that the flux per plaquette is
of  the form $p/q$ where $p,q$  are co-prime integers. 
We consider fermion densities corresponding to filled bands and address following two questions:  (i) How
common are translation symmetry breaking transitions and how often are they
accompanied by the change in the Hall conductivity ? (ii) Do they destroy the fractal 
structure ?

The rest of this paper  is  organized as follows: In Sec.~\ref{sec:II}, we
discuss the model and the phase transitions in the system due to the
interactions.  Sec.~\ref{sec:III}, gives a brief review of the non-interacting
Hofstadter butterfly in the honeycomb lattice and  describes the self
similarity of  the fractal structure of the butterfly. The effect of
interactions  on the Hofstadter butterfly is   described in Sec.~\ref{sec:IV}.
In Sec.~\ref{sec:V}, the  Landau fan diagram for both the  non-interacting and
interacting  cases  are discussed and compared. Finally, we conclude  in
Sec.~\ref{sec:VI}.

\section{\label{sec:II} Model and phase transitions}

The  model  we  consider  is spinless fermions on the honeycomb lattice  in the
Hofstadter regime with nearest neighbor hopping and nearest neighbor
interaction.  The  Hamiltonian is 

\begin{equation} 
\label{ham1} H=-t\sum_{\langle ij\rangle}
\left(c^\dagger_i e^{i\frac{e}{\hbar}A_{\langle ij\rangle}}c_j
+h.c\right) +V\sum_{\langle ij\rangle}n_in_j, 
\end{equation} 
where $c_{i}~(c_{i}^{\dag})$ is the annihilation (creation) operator for
electrons at site $i$ on the honeycomb lattice, $n_{i}$ is the number density
operator and
$t$ is the nearest  neighbor hopping parameter and 
$V$ is the nearest
neighbor interaction strength.
We consider $t=1$ and $V$ is in units of $t$.
$A_{\langle ij\rangle}$ are the gauge fields on the
nearest neighbor links such that the magnetic flux passing through
each plaquette is $\phi=\frac{p}{q}\frac{h}{e}$ where $p,~q$ are co-prime integers 
with $q=3,\cdots 20$ and $p<q$. We refer different values of flux per plaquette as different systems. For the range of $q$
considered in this paper, there are $126$ flux values in total and thus $126$ systems.
For each of these systems,
we solve the interacting problem for the filled band cases only. Hence, for
a particular interaction strength,
we have $3484$ cases in total.

\begin{figure}[hbtp]
\begin{center}
\includegraphics[scale=0.45]{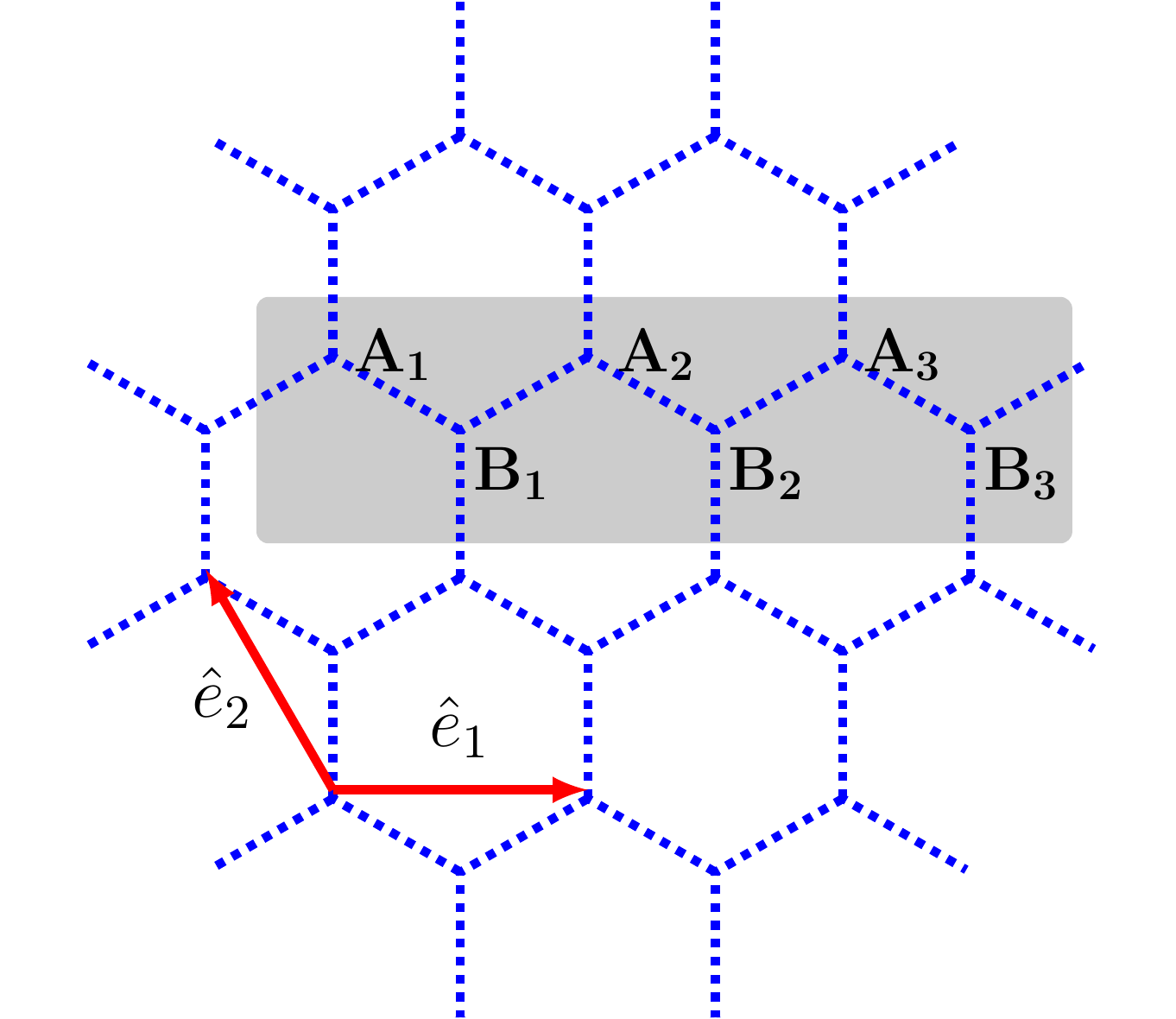}
\caption{\label{fig:0}(Color Online) Honeycomb lattice in magnetic
  field with flux $\phi=1/3$ passing through each plaquette. 
  $A$ and $B$ are the two sublattices. $\hat {e_{1}}$ and $\hat {e_{2}}$ represent the basis vectors of the
  lattice. The gray portion shows the magnetic unit cell choice considered in this paper. $A_1, B_1, A_2, 
B_2, A_3, B_3$ are sublattices of this magnetic unit cell.}
\end{center}
\end{figure}

The Hamiltonian is invariant under magnetic translations $\tau_1$ and $\tau_2$
which are along $\hat e_1$ and $\hat e_2$ directions respectively.
$\tau_1\tau_2\tau_1^{-1}\tau_2^{-1}=e^{i\frac{2\pi}{q}}\Rightarrow
[\tau^q_1,\tau_2]=0$. We choose the magnetic unit cell to be $q$ adjoining
original unit cells along the $\hat e_1$ direction as shown in Fig.~\ref{fig:0}
for $q=3$.  Each magnetic unit cell contains $2q$ sites. Other symmetries of the
system are 6-fold rotations about the centers of
the hexagons, 3-fold rotations about the sites and 2-fold rotations (inversion)
about the centers of the links. At half filling, the system also has
particle-hole (chiral) symmetry, $c_i\rightarrow (-1)^{p_i}c^\dagger_i$, where
$p_i=0$ for $i$ belonging to one of the sublattices and $p_i=1$ for the other.

The Brillouin zone is the set of wave vectors $\vec k=k_1\vec G_1+k_2\vec G_2$, where $\vec
G_{1,2}$ are the reciprocal lattice vectors of the underlying triangular
lattice with $-\pi/q\le k_1\le \pi/q$ and $-\pi\le k_2\le\pi$.

To solve this interacting problem, we use mean field approximation discussed 
in  our previous work \cite{archana2016},
\begin{align}
\nonumber
n_in_j &\approx \left(\Delta_ic^\dagger_jc_j
+\Delta_jc^\dagger_ic_i\right)
-\chi_{\langle ij\rangle}c^\dagger_ic_j
-\chi_{\langle ij\rangle}^*c^\dagger_jc_i\\
&-\frac{1}{V}\left(\Delta_i^2+\Delta_j^2
-\vert\chi_{\langle ij\rangle}\vert^2\right),\\
\frac{1}{V}\chi_{\langle ij\rangle}&=\langle c^\dagger_jc_i\rangle_{MF},
~~~~~~
\frac{1}{V}\Delta_i=\sum_{j(i)}\langle c^\dagger_jc_j\rangle_{MF},
\label{sce}
\end{align}
where $j(i)$ denotes all the nearest neighbors of $i$. The self consistency
equations, Eq.~\eqref{sce}, have to be solved keeping the number density 
fixed. We solve them numerically for interaction strength $V=1,~2,~4$. 
For each flux value $p/q$, there are  $3q$ complex  bond order  parameters 
and  $2q$  real charge order parameters. 

The choice of the magnetic unit  cell is not unique. For the non-interacting
case, when the translation symmetry is not broken, the  choice of the unit cell is
irrelevant. For the interacting  case, the choice of unit  cell matters.  It
determines the pattern of the translation symmetry breaking.  With increasing
$q$,  the number of distinct magnetic unit cell choices increase and it is not
feasible to solve all the possibilities numerically.  We restrict ourselves to
the linear magnetic unit  cell choice, as shown in Fig.~\ref{fig:0} for $p/q=1/3$.
There could be phases with other patterns of translation symmetry breaking with lower
energy and they could appear at lower values of the interaction strength.  Our
analysis thus underestimates the effects of the interactions.

We work with a lattice of $30\times30$ magnetic unit cells and a fixed number
of particles corresponding to a particular band filling. The self consistency
equations are  solved for filled bands till half filling.
The self consistency solutions for the upper half filling cases is same as the
lower half due to particle  hole symmetry. 

We solve the mean field Hamiltonian for these cases for the interaction
strength $V=1,~2,~4$. We observe that for $V=1$, $41\%$ of the cases show phase transitions
and $31\%$ of the cases show topological transitions, for $V=2$, $70\%$ of the cases show
phase transitions and $58\%$ of the cases show topological transitions  and for
$V=4$, $84\%$ of the cases show phase transitions and $71\%$ of the 
cases have topological transitions. Hence, we see that even for interaction strength $V=1$
significant number of cases show phase transitions and topological phase transitions which
increases with increase in the interaction strength.

\begin{figure}[h!]
\centering
\subfloat[]{\label{fig:1a}\includegraphics[scale=0.75,trim=15mm 0mm 0mm 0mm,clip]{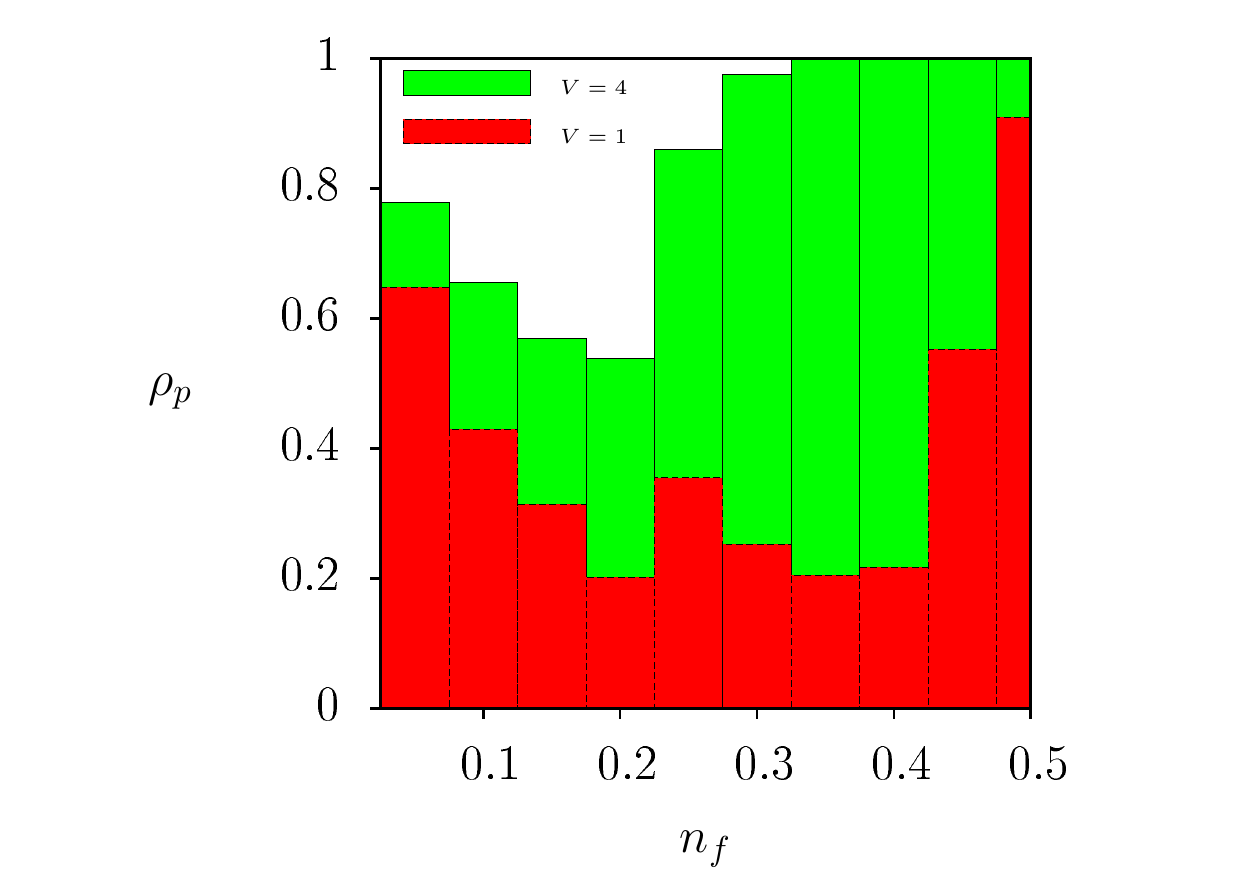}}\\
\subfloat[]{\label{fig:1b}\includegraphics[scale=0.75,trim=15mm 0mm 0mm 0mm,clip]{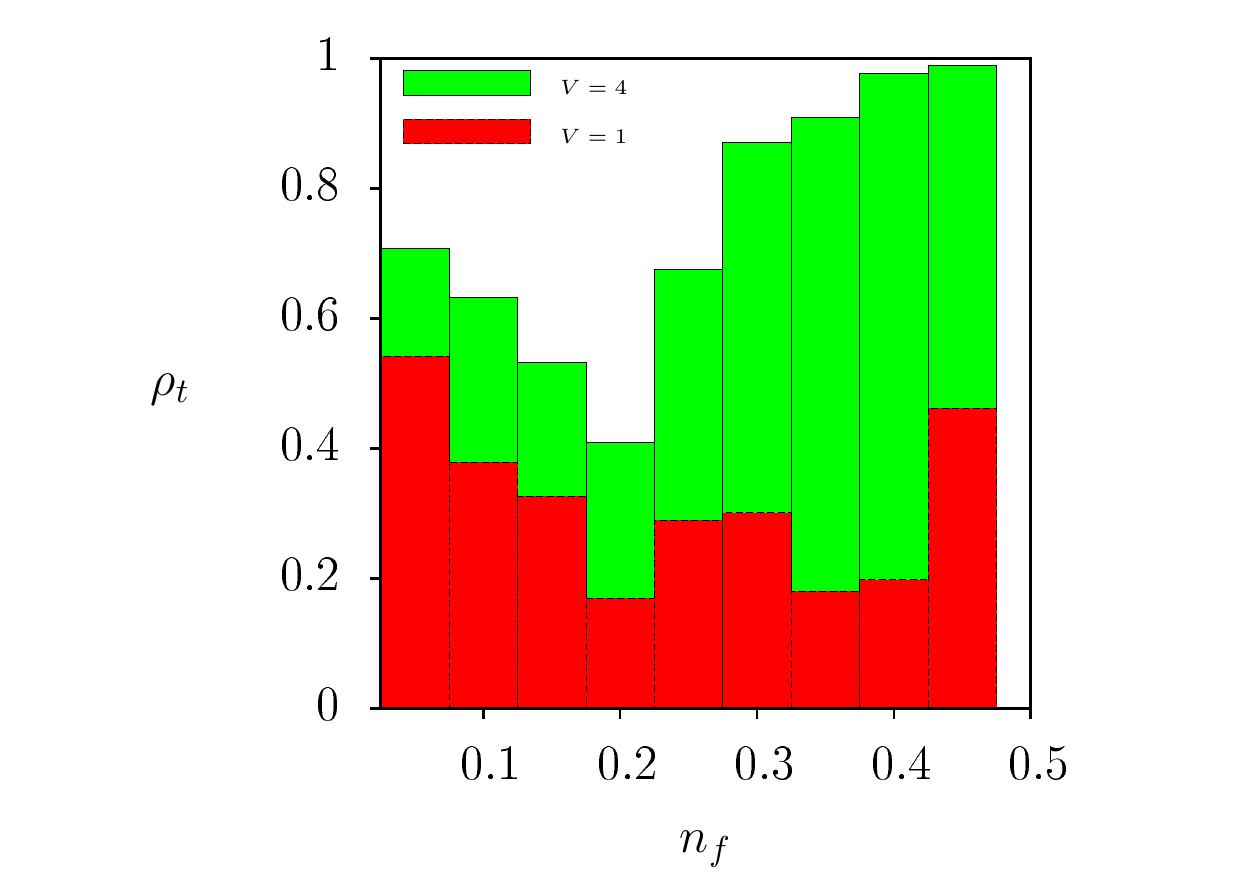}}
\caption{\label{fig:1}(Color Online) Bar plot for (a) probability of the phase transitions ($\rho_p$) vs filling fraction $n_f$ for 
$V=1$ and $V=4$ and (b) probability of the topological phase transitions $\rho_t$ vs $n_f$ for $V=1$ and $V=4$.}
\end{figure}

Fig.\ref{fig:1} shows the probability of
phase transitions, $\rho_p$,  and topological phase transitions, $\rho_t$, as a
function of the filling fraction, $n_f$. $\rho_{p(t)}$ is defined as the number
of systems with filling between $n_f$ and $n_f+dn_f$ that show phase
transitions (topological transitions) divided by the total number of systems
with filling between $n_f$ and $n_f+dn_f$. In Fig.\ref{fig:1}, we have taken
$dn_f= 0.05$.

Fig.\ref{fig:1} shows that the number of phase transitions peak at half filling
and at the dilute limit. As expected the number increases with 
increase in the interaction strength. The peak near half-filling seems intuitively
reasonable, since the inter-particle distance decreases with increasing 
particle density and so the effect of the nearest neighbor interaction 
increases.  However, by the above reasoning, there should be minimum number of phase 
transitions in the dilute limit, quite  contrary to what we see in 
Fig.\ref{fig:1}.

\begin{figure}[h!]
\label{fig:01a}\includegraphics[scale=0.13,trim=15mm 10mm 10mm 10mm,clip]{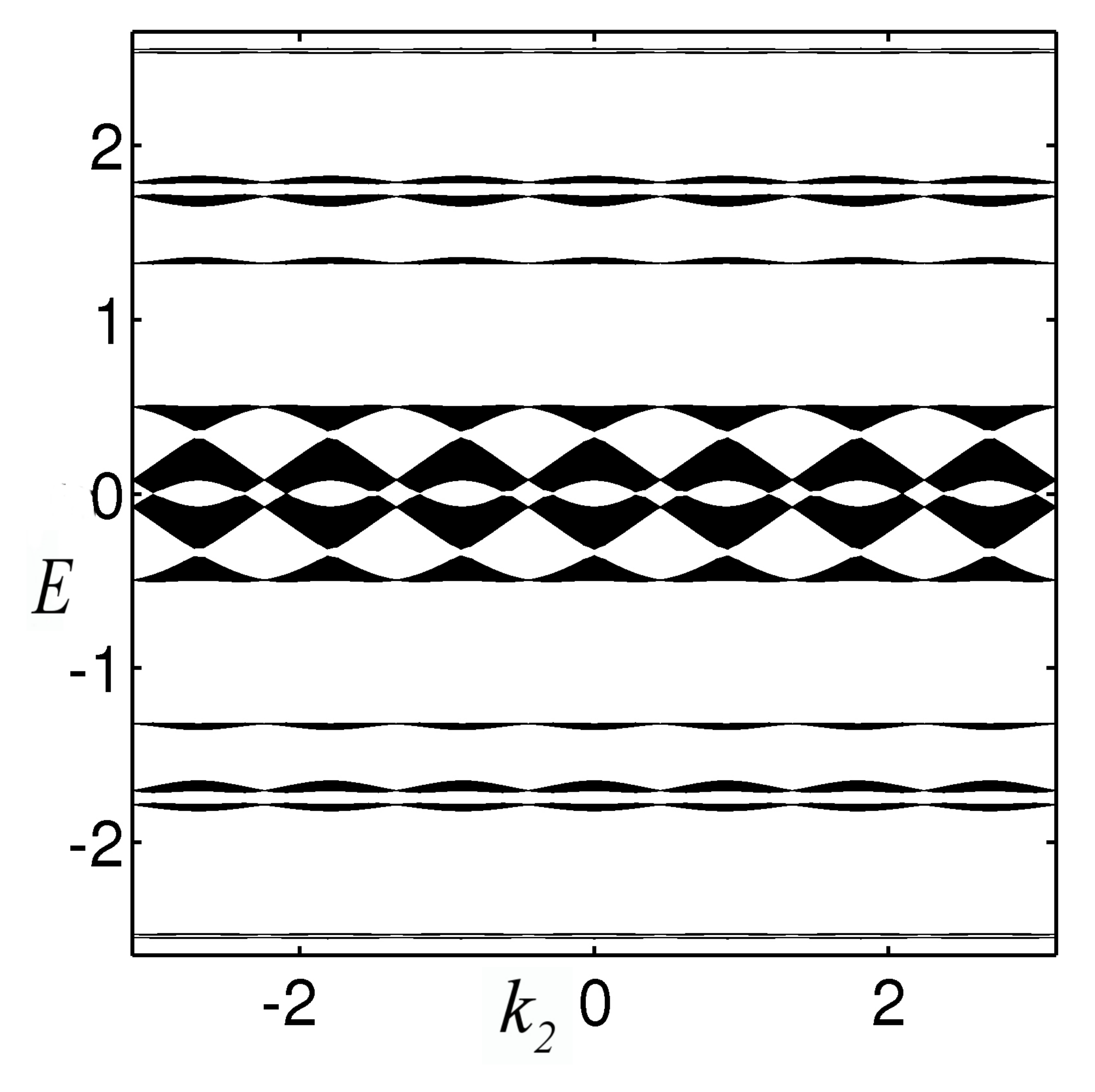}
\caption{\label{fig:01}(Color Online) Non-interacting energy band diagram for $p/q=2/7$. 
In this figure, in the Fermi regime, the lowest two bands are very close to each other and bunch together.
Similarly in the Dirac regime, at half filling, the four energy bands bunch together.}
\end{figure}

The answer to this puzzle comes from examining the energy bands of the
non-interacting system. We observe that for flux per plaquette $p/q$, at low filling, $p$
bands come close to each other and the energy gap between these bands decreases
with the increase in $q$. 
Since the band gaps
tend to become low in these regions for the systems with larger $q$, it is
easier for the interaction to mix the bands leading to the transitions.
Near half filling $2p$ bands bunch up and tend to get
degenerate at large $q$.  Fig.~\ref{fig:01} illustrates this for flux value
$2/7$. In the Fermi regime,  the energy gap between the lowest two bands in
Fig.~\ref{fig:01} is very small and are bunched together. These gaps  become negligible as $q$
increases. Similarly, the four energy bands near the half filling, the Dirac regime, bunch together as can be
seen from the Fig.~\ref{fig:01}. On increasing $q$, these bands become $2p$ degenerate.	

This phenomenon of bands bunching as $q$ increases can be understood by
examining two extreme limits of this problem: (i) Hofstadter regime (small $q$)
(ii) Weak field limit ($q\rightarrow \infty$). At small $q$, we typically have
$2q$ well separated bands, each of them contributing a particle density of
$1/q$ per unit cell, when completely filled. 

At large $q$, we can analyze the system in the continuum limit, separately for
the dilute limit and near half-filling. In the dilute limit, the system behaves
like a single species of non-relativistic fermions in a magnetic field.  The
spectrum in this regime, dubbed as the Fermi regime by Hatsugai et. al.
\cite{hatsugai2006}, consists of Landau levels each contributing a particle
density of $p/q$ per unit cell, when completely filled.  Thus we may expect $p$
of the bands to become degenerate in the weak field limit, consistent with the
bunching that we observe.

Near half-filling is the so called Dirac regime \cite{hatsugai2006}. Here the
system behaves like two species of Dirac quasiparticles. The spectrum consists
of relativistic Landau levels. Since there are two species, each Landau level
has a particle density of $2p/q$. Thus in this regime we expect a bunching of
$2p$ bands with increasing $q$, which form the degenerate Landau level in the
$q\rightarrow\infty$ limit.

From the above argument, we also expect the Chern number of the bunch of $p$
bands to sum up to 1 in the dilute limit and that of the bunch of $2p$ bands to
sum up to 2 near half-filling. We have computed the Chern numbers numerically
and have found that this is indeed true.

\section{\label{sec:III}Self similar structure of the non-interacting Hofstadter butterfly in honeycomb lattice}

Before describing the effect  of interactions on the Hofstadter butterfly, we
briefly review its fractal structure for the non-interacting honeycomb lattice.
In this section, we also show that the range of $q\leq20$ we are considering
is large enough to see the self similarity to first order, namely 
the first step of the recursion.

\begin{figure}[h!]
\centering
\includegraphics[scale=0.9,trim=15mm 0mm 0mm 0mm,clip]{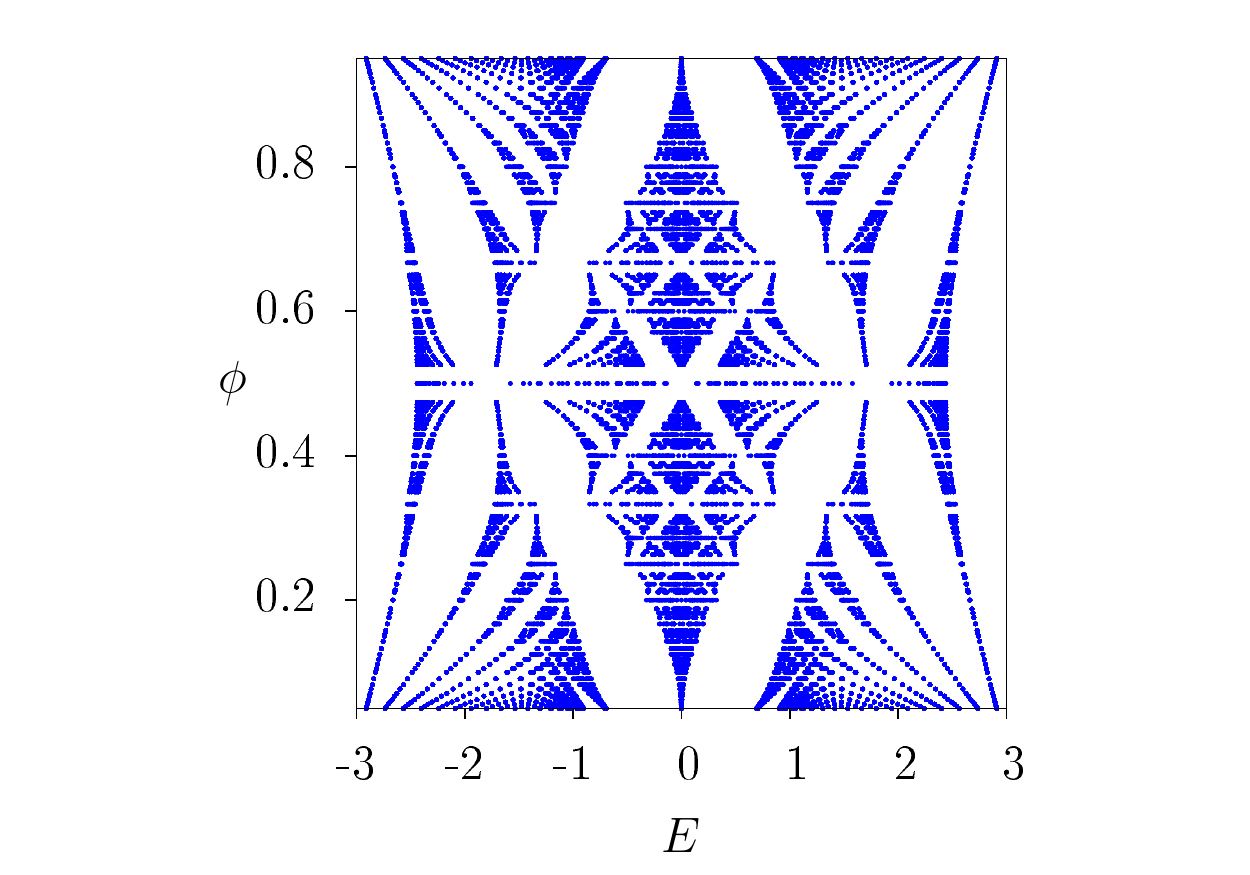}
\caption{\label{fig:2}(Color Online) Hofstadter butterfly for the honeycomb lattice.
Here the x-axis represents the single particle energy $E$ and y-axis is the magnetic flux per plaquette $\phi$ 
of form $p/q$.  In this plot $q\leq 20$.}
\end{figure}

The Hofstadter butterfly for the  non-interacting honeycomb lattice
\cite{rammal1985,gumbs1997,goldman2009,claro1979} is  shown in Fig.~\ref{fig:2}.  
This plot is self similar, in the sense that there is a 
mapping of the whole plot into a block in the plot which is described in detail in this section. 
When this block is suitably
scaled and rotated, it reproduces the original plot. We describe the recursion below.

\begin{figure*}[h!]
\centering
\subfloat[]{\label{fig:4a}\includegraphics[scale=0.72,trim=16mm 0mm 2cm 3mm,clip]{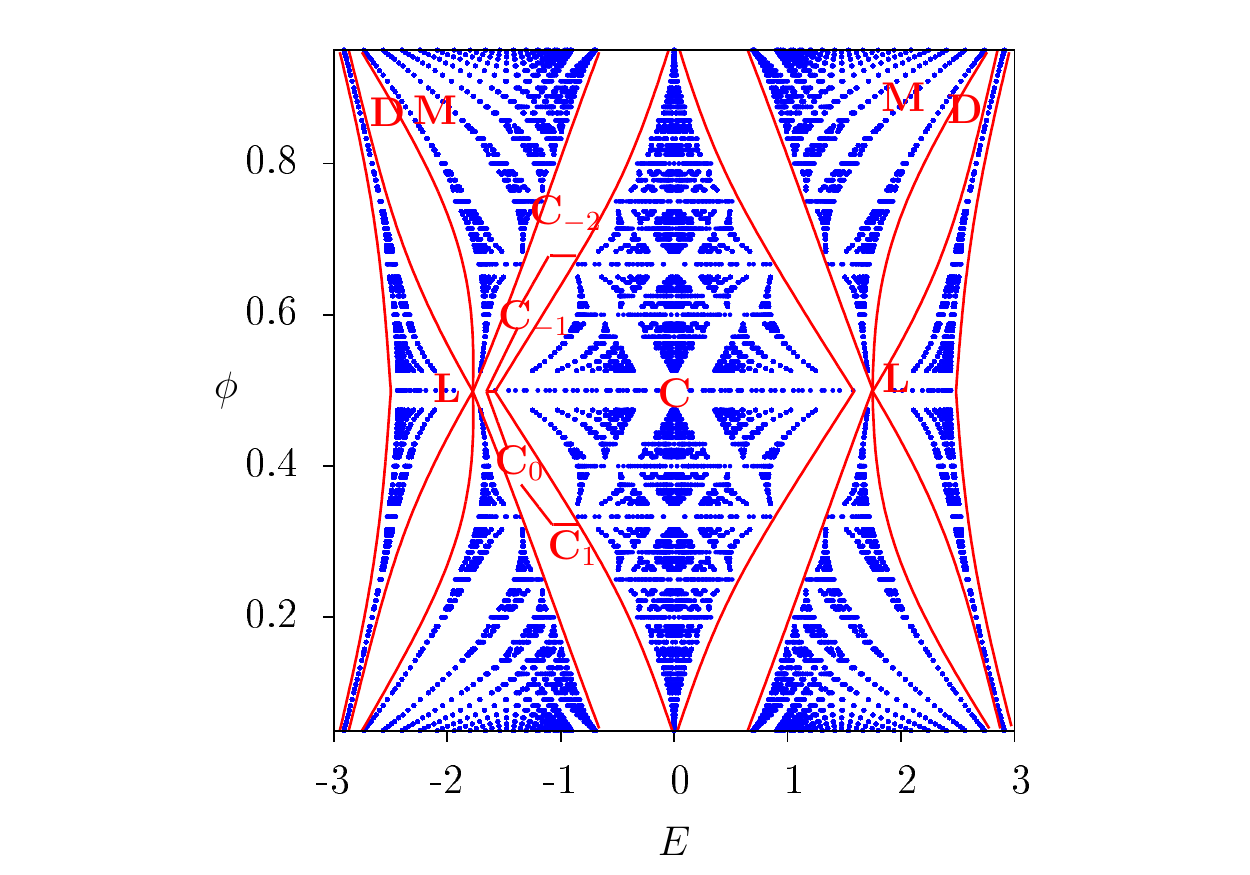}}\hfil
\subfloat[]{\label{fig:4b}\includegraphics[scale=0.72,trim=15mm 0mm 2cm 3mm,clip]{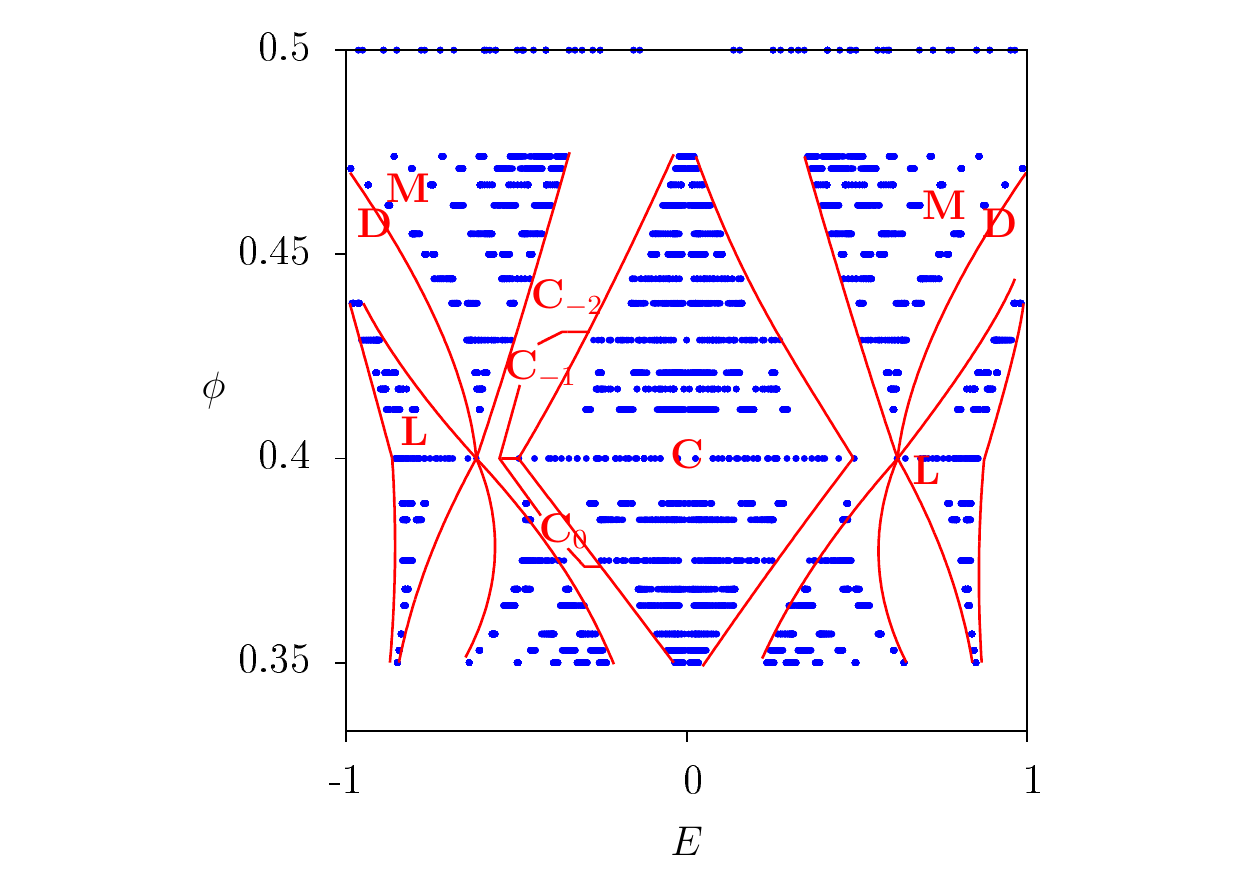}}
\caption{\label{fig:4.4} (Color Online) Hofstadter butterfly for the  honeycomb lattice where the energy spectrum 
is plotted for the flux per plaquette in the range(a) $(1/20,19/20)$  and energy range $[-3,3]$ and (b) $(1/3,1/2)$
and energy range $[-1,1]$. Fig.~\ref{fig:4b}
is the plot of the Hofstadter butterfly in $C_0$ sub-block.}
\end{figure*}

First we construct the so called `skeleton' of the Hofstadter butterfly.  We
define $\phi$ of the form $1/q$ and $1-1/q$ for $q\geq2$ to be `pure cases'.
The skeleton is formed from the set of curves constructed as described below:  (a) Connect
the outer edges of $q^{th}$ and $(q+1)^{th}$  band of neighboring pure cases
for $q\leq2$.  This forms a  huge box which we denote as the $C$ block.
We divide this $C$ block into sub-blocks. These sub-blocks are the portions of the $C$ block  between $q+1^{th}$ and $q^{th}$
flux values. We label these sub-blocks as $\cdots,~C_{-1},~C_0, ~C_1,~\cdots$ as shown in Fig. \ref{fig:4a}.
(b) Connect the right  outer edges of $(q-1)^{th}$  band 
and the left outer edges of the lowest band of
neighboring pure cases for $q\leq2$.  This forms a  huge box which we denote
as the $D$ block. The $D$ block is further divided into $L$ and
$M$ sub-blocks.  Connecting the right outer  edges of the lowest band of the
neighboring pure cases form the $L$ sub-block and connecting the left outer
edge of the second lowest band and the right outer  edge of the $(q-1)^{th}$
bands  of the neighboring pure cases form the $M$ sub-block. This construction
is shown in Fig.~\ref{fig:4a}.

The whole plot in a compressed form with some rotation is present inside each of the $C$
sub-blocks \cite{rhim2012}. This statement is quantified using a recursive relation from 
the original plot to the $C$ sub-blocks \cite{rhim2012}.
In each of these sub-blocks, there is a local variable, $\phi'$, for flux per plaquette defined in terms of 
the variable $\phi$ representing the flux per plaquette of the original plot.
For $\phi\leq 1/2$ we define $N$ as $N\equiv[1/\phi]$. $[x]$ stands 
for the greatest integer less than or equal to $x$.
The recursive relation between $\phi$ and $\phi'$  is given by \cite{rhim2012}
\begin{eqnarray}
\phi&=&\frac{1}{N+\phi'},~\phi\leq \frac{1}{2}\nonumber\\
1-\phi&=&\frac{1}{N+\phi'},~\phi\geq\frac{1}{2}
\label{req}
\end{eqnarray}
Thus the local variable $\phi'$ has  values in  $[0,1]$ like the flux  in  
the original plot. Self-similarity requires that the number of energy bands
and gaps at a value of $\phi$ in the original plot is the same as the number 
of energy bands and gaps at $\phi'$ in each sub-block.
\begin{figure*}[h!]
\centering
\subfloat[]{\label{fig:6a}\includegraphics[scale=0.72,trim=15mm 0mm 2cm 2mm,clip]{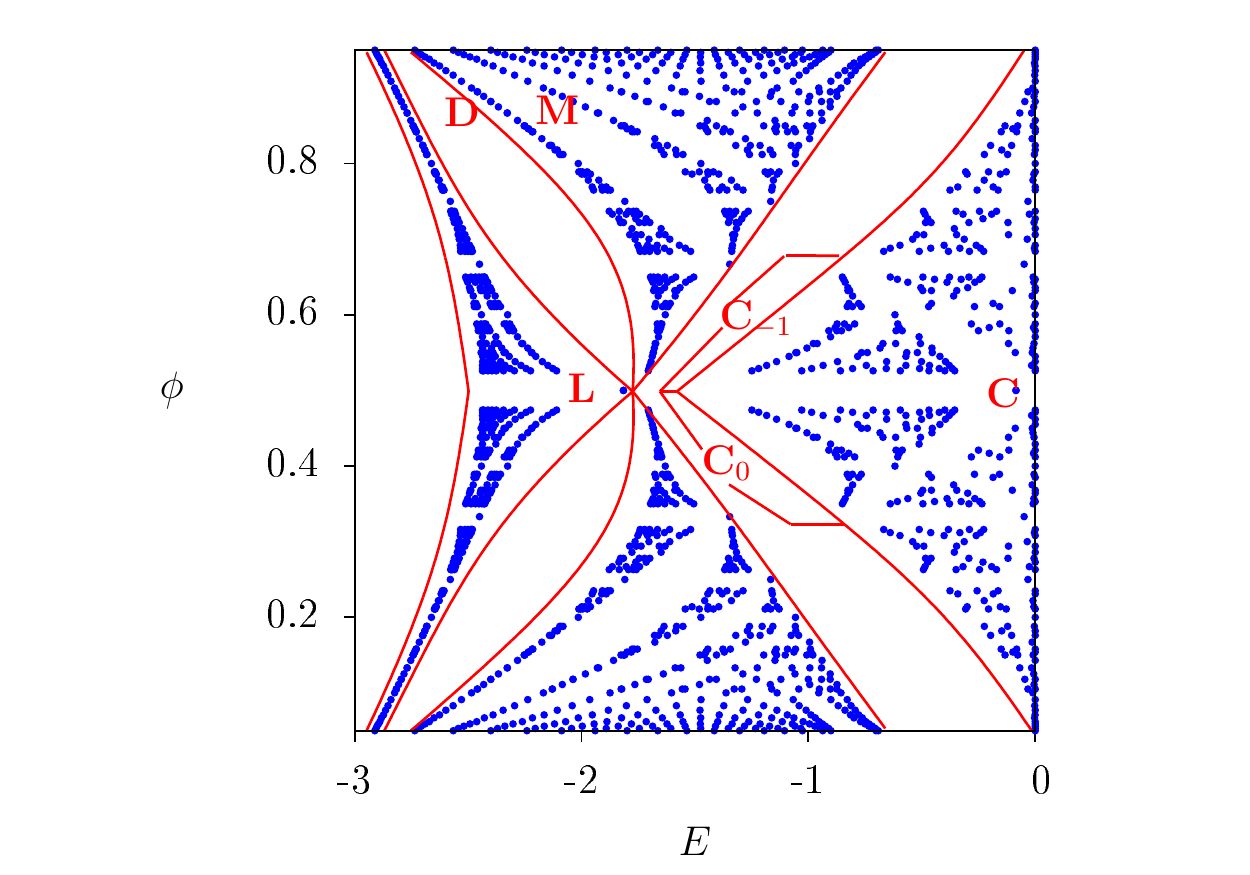}}\hfil
\subfloat[]{\label{fig:6b}\includegraphics[scale=0.72,trim=15mm 0mm 2cm 2mm,clip]{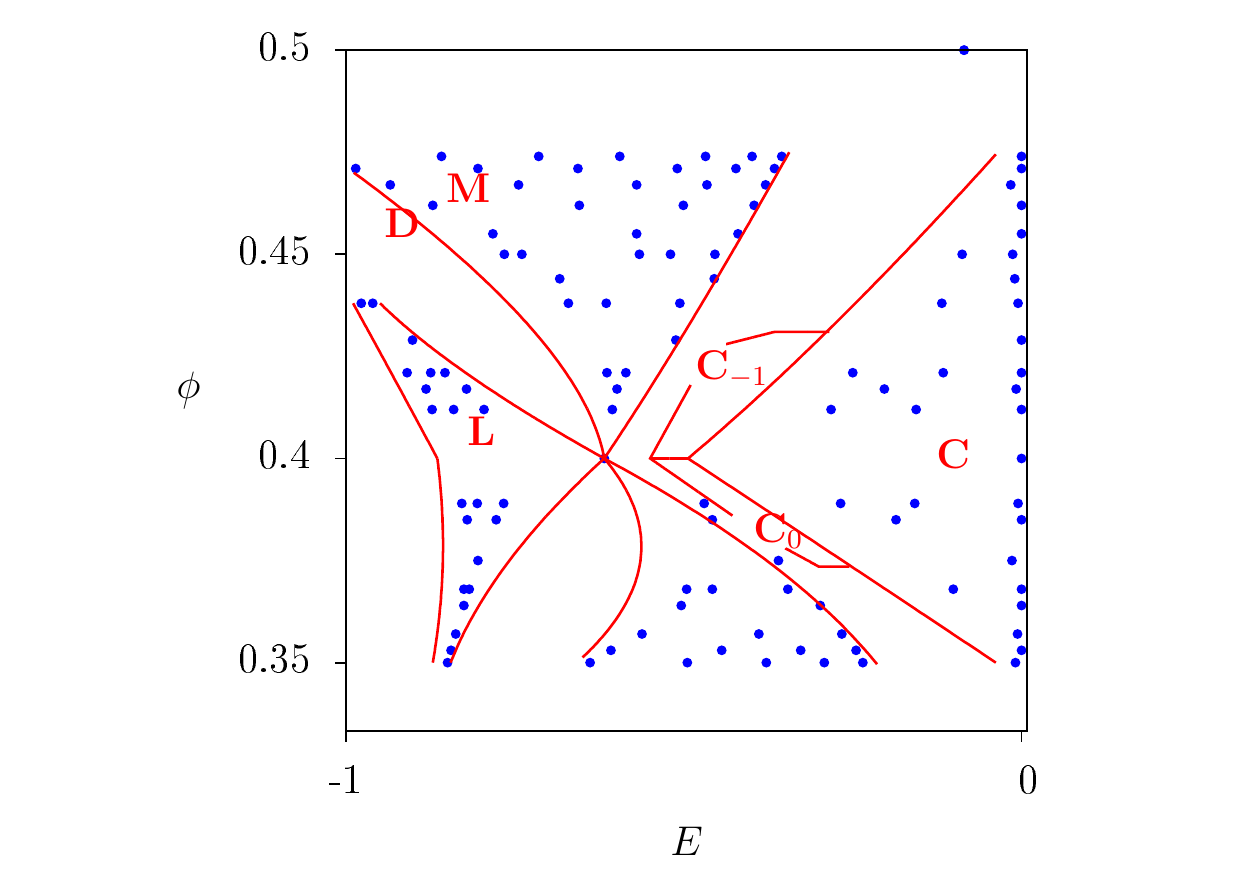}}
\caption{(Color Online) Hofstadter butterfly for the honeycomb lattice in the  absence of interaction plotted by taking
the maximum energies of each band as the x-axis instead  of the whole energy  spectrum for
(a) flux in the range $(1/20,19/20)$ and  the energy in the range [-3,0] 
(b) flux in  the range $(1/3,1/2)$ and energy in range [-1,0]. The  maximum energy is plotted till half-filling.}
\end{figure*}


We now show that $q$ in the range $[2,20]$
is enough to verify the first step of the  recursion
. In Fig.~\ref{fig:4a}, the $C$ block has $36$
sub-blocks for the considered range of $q$. 
 Each value of $\phi$ in $C_0$ can be mapped to a value of $\phi'$ using Eq. \eqref{req}.
Similar to $\phi$, we can define `pure cases' in $\phi'$.
For example: 
$\phi=2/5\Leftrightarrow\phi'=1/2$; in this case 
 $C_0$ sub-block has $4$ bands which is the same number of bands for $\phi=1/2$
in original plot, 
$\phi=3/7\Leftrightarrow\phi'=1/3$; here $C_0$ sub-block has $6$ bands which is the same number of bands for $\phi=1/3$
in original plots 
 as seen in Fig. \ref{fig:4.4}. 
 In general, the number of bands for a particular value of $\phi'$ in $C_0$ sub-block is same as that of $\phi=\phi'$
in the original plot.
 Using the same method of constructing blocks in the original unit cell, as described before, we
construct the secondary $C$ and $D$ blocks in the $C_0$ sub-block. This secondary $C$ block is further divided into $12$
sub-sub-blocks some of which are shown in Fig.~\ref{fig:4b}. We verify that each of these secondary blocks have the same number
of energy bands and gaps and are arranged in similar  fashion for a  particular value of $\phi'$ as the original plot
for the same value of $\phi$. Hence, the energy spectrum of the original plot is seen to be
repeating in the  sub-block and the self-similarity criterion is satisfied upto first order for the
 $q\in[2,20]$.

\section{\label{sec:IV}Effect of Interactions}

The Hofstadter butterfly described in the previous section is defined  for the
non-interacting case in terms of the single particle energies and the single
particle gaps. However, single particle energies are  not defined for the
interacting system. In presence of interactions, the well defined quantities
are the many body ground state at each electron densities and the chemical
potential at these number densities. In the non-interacting system, the
chemical potential is the highest occupied single particle energy. Hence,
in the interacting case, it is natural to plot the magnetic
flux per plaquette with  respect to the chemical potential for different
fillings.

We solve this interacting problem using mean field theory. 
The mean field Hamiltonian, $H_{MF}$. can be defined as 
\begin{equation}
H_{MF}=\int_k c^\dagger(k)h_{MF}(k)c(k)
\end{equation}
where $c(k)$ is a $2q$ component vector with components $c_{k \alpha},~
\alpha=1,\dots 2q$. $c_{ \alpha}$ is the Fourier transform of the fermion operators $c_{i\alpha}$
and is  defined as,
\begin{equation}
c_{k\alpha}=\frac{1}{\sqrt N}\sum_n~e^{i\vec k\cdot\vec R_n}c_{i\alpha}
\end{equation}
where $c_{i\alpha}$ is the fermionic operator that annihilates an  electron at
unit cell $i$  and sublattice  index  $\alpha$, $N$ is the total number of sites and $\vec k\in$ Brillioun zone.
$h_{MF}(k)$  is the single particle mean field Hamiltonian in the momentum
space which is a $2q\times 2q$ matrix.
The spectrum of $h_{MF}(k)$ is given by the eigenvalue equation
\begin{equation}
h_{MF}(k)u^m(k)=\epsilon^m(k)u^m(k)
\end{equation} 
where $\epsilon^m(k)$ is  the single particle energy for the $m^{\rm th}$ band
and $u^{m}(k)$ is its eigenfunction.  $c^\dagger_{m'k}=\sum_{\alpha}
c^\dagger_{k \alpha}u^{m'}_{\alpha}(k)$ is the fermionic operator that creates
a fermion in the single particle state with quasi momentum vector $k\in BZ$ in
the $m'^{\rm th}$ band.

For $m$ filled bands, the mean field ground state can be written as 
$$\vert m\rangle=\prod_{m'=1}^m\prod_{k\in BZ}c^\dagger_{m'k}\vert 0\rangle$$

where $\vert 0\rangle$ is the vacuum state.  The ground state energy is the sum
of the single particle energies $\epsilon^{m}_{k}$.  If the state,
$c^\dagger_{m+1 k_0}\vert m\rangle$ is a good approximation of the single
quasi-particle state, then the gap is the lowest single particle energy in the
$m+1^{\rm th}$ band, $\epsilon^{m+1}_{k_0}$. 

While this is often the case, it is not always so. For example, in the case
of quantum Hall skyrmions \cite{sondhi1993,brey1997}, the order parameter
deforms locally when one electron is added to the system and the gap can 
get reduced by a factor of 2. When we solve our mean field equations for 
a dilute density of fermions in the $m+1^{\rm th}$ band, we find that even 
for very small filling the mean
field Hamiltonian and consequently the ground state changes significantly i.e
$\vert m+\delta\rangle$ is quite different from $\vert m\rangle$ even for small
$\delta$. We feel that this may imply that the order parameters
deform locally when an extra particle is added and hence that the mean field
gap is not reliable. Exploring this issue in detail is ongoing work.

In this work, we consider only the filled band cases where the mean field 
theory is reliable. While we do not
know the single particle energy gaps, nevertheless, we can take the band edge
to be the single particle energy of the highest occupied level (which is the
chemical potential). Hence, to investigate the effects of the interactions on
the Hofstadter butterfly, we first plot the magnetic flux per plaquette versus
the maximum energy of the band for the non-interacting case and show that this
plot also shows the self-similar structure. We then compare it with the
interacting case.

Fig.~\ref{fig:6a}  is the plot  for flux per  plaquette  versus the maximum energy of  each band for the
non-interacting case. The plot is restricted  to  half-filling here. 
\begin{figure*}[h!]
\centering
\subfloat[]{\label{fig:7a}\includegraphics[scale=0.72,trim=15mm 0mm 2cm 0.2cm,clip]{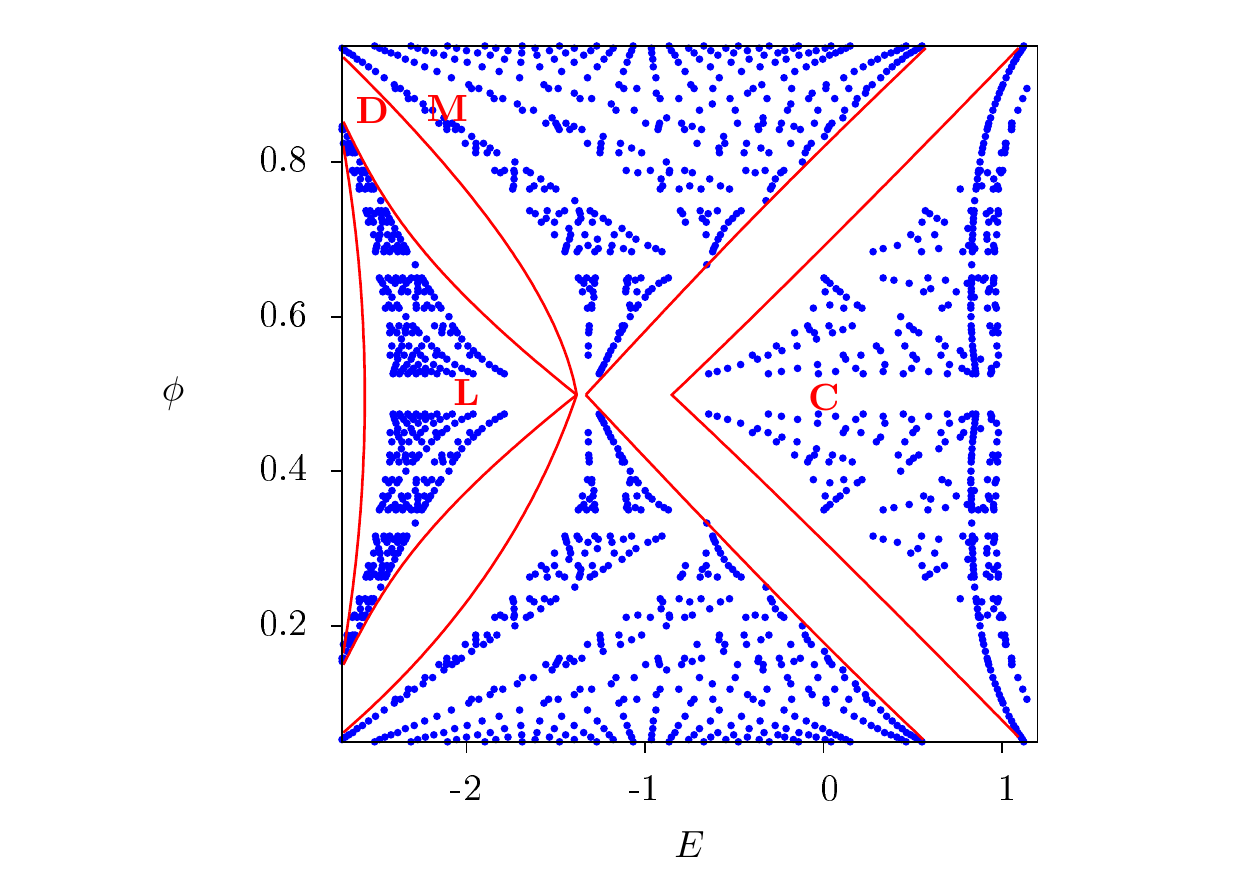}}\hfil
\subfloat[]{\label{fig:7b}\includegraphics[scale=0.72,trim=15mm 0mm 2cm 0.2cm,clip]{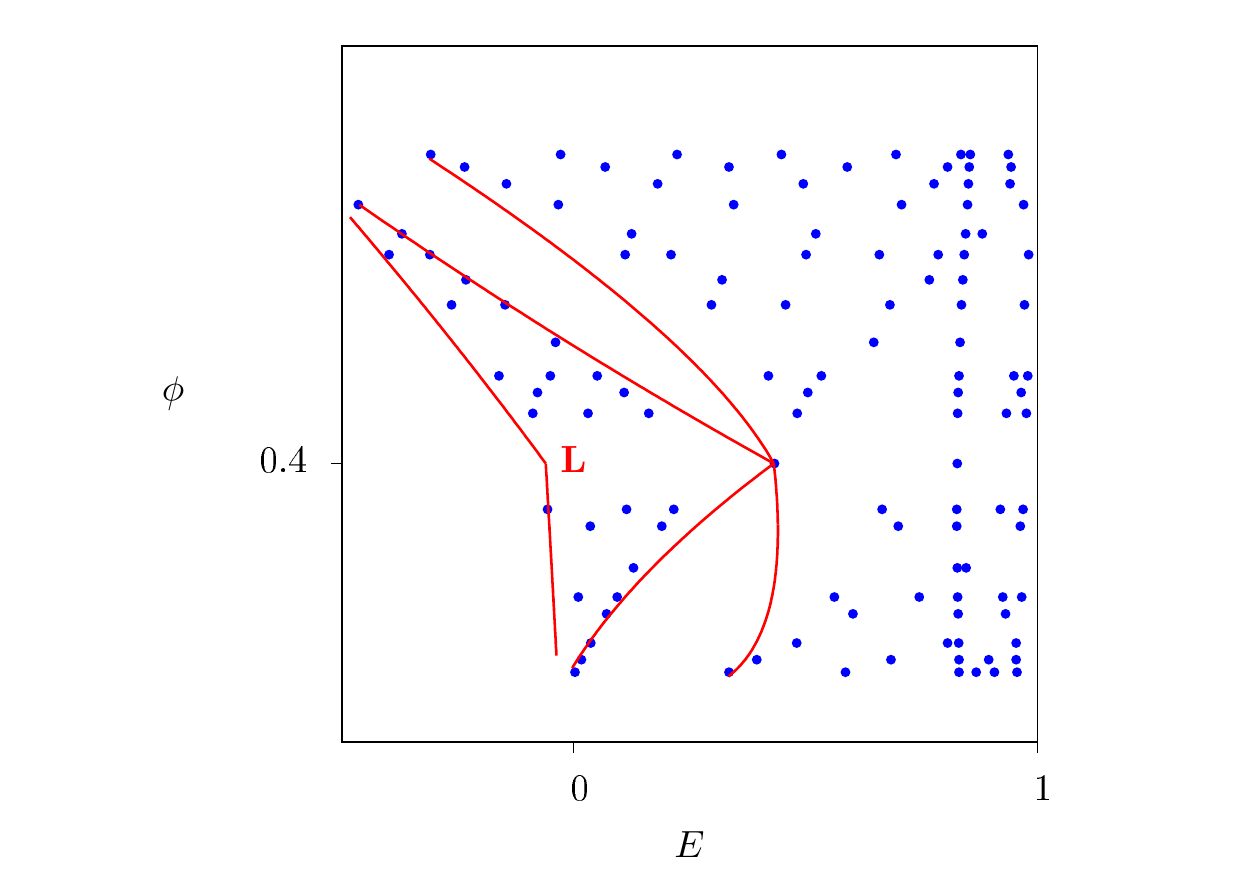}}\\
\subfloat[]{\label{fig:7c}\includegraphics[scale=0.72,trim=15mm 0mm 2cm 0.2cm,clip]{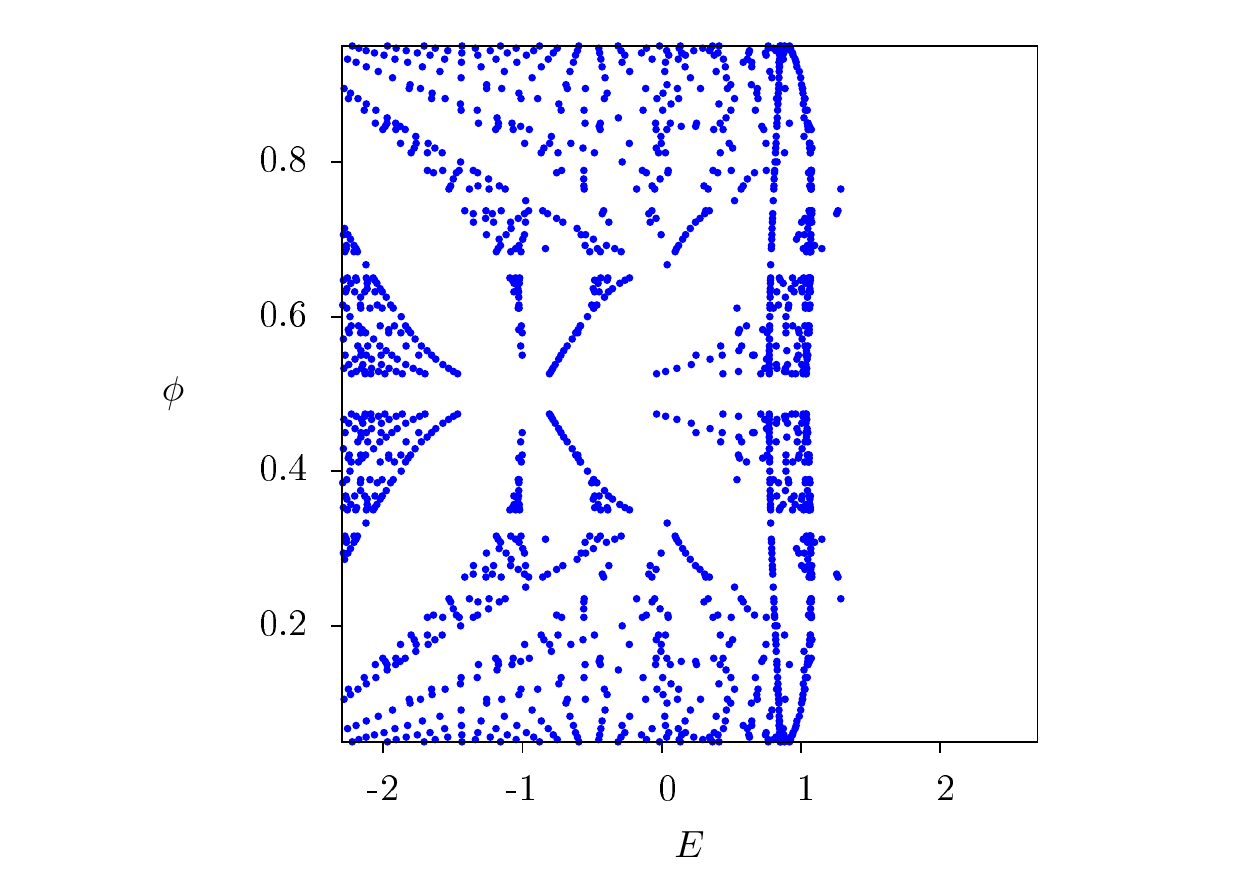}}\hfil
\subfloat[]{\label{fig:7d}\includegraphics[scale=0.72,trim=15mm 0mm 2cm 0.2cm,clip]{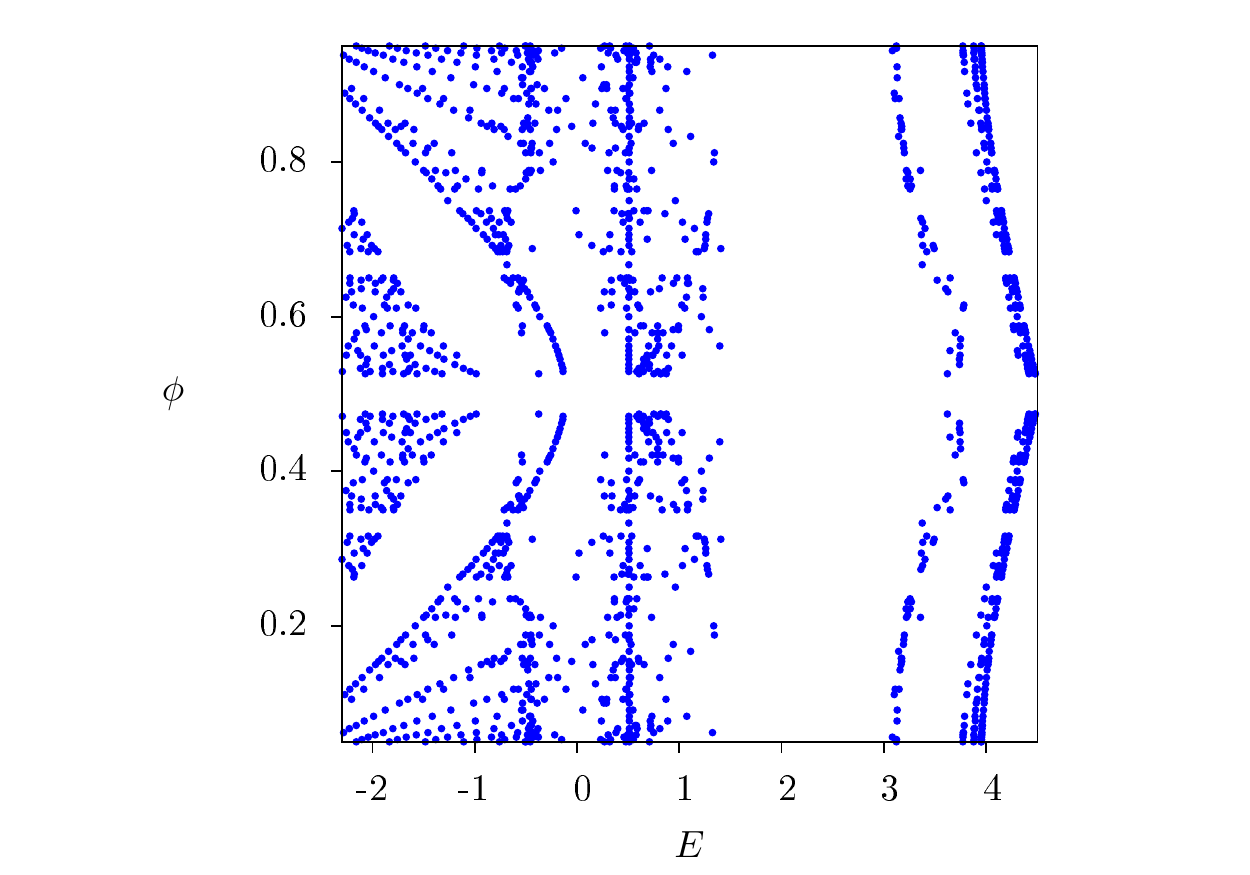}}
\caption{\label{fig:7} (Color Online) Hofstadter butterfly for the  honeycomb lattice in the presence of interaction plotted by taking
the maximum energy as the x-axis for the
flux in the range $(1/20,19/20)$ for (a) $V=1$, (c) $V=2$ and (d)  $V=4$. 
(b)Hofstadter butterfly for the  honeycomb lattice in the presence of interaction plotted by taking
the maximum energy as the x-axis for the
flux in the range $(1/3,1/2)$ for $V=1$. The maximum energy is plotted till half-filling }
\end{figure*}
As seen in Fig.~\ref{fig:6a}, the plot is divided into  blocks $C$ and $D$ (including $M$  and $L$ blocks).
The $C$ block is further  divided into sub-blocks where  this pattern repeats itself as seen in Fig.~\ref{fig:6b}.
In this case, the self similarity
criterion is satisfied by verifying that the number of energy points for a particular value of $\phi'$ in $C_0$
sub-block is same as that of $\phi$ = $\phi'$ in the original plot. Fig.~\ref{fig:6b} is the 
plot of the  Hofstadter butterfly in the absence of interaction for the flux in the range $(1/3,1/2)$ i.e. the $C_0$ sub-block of
the $C$ block. 
For the range of $q$ considered in this paper, we verify that each of
these secondary blocks have the same number of energy points and are arranged in similar fashion for
a particular value of $\phi'$ as the original plot for the same value of $\phi$. Thus, we show that the recursive pattern
is present even for plot of magnetic flux per plaquette versus the maximum energies of the bands upto first step.

Now we study the effect of the interaction on Fig.~\ref{fig:6a}.  As described
in our previous work \cite{archana2016}, there  is always a  scaling solution
for  this  interacting problem which satisfies  the  self consistency
equations. The state corresponding to this solution is the same as the
noninteracting case. However, the strength of the hopping parameters get scaled
as $t\rightarrow\lambda t$ and the single particle energies get scaled as $E_0(t)\rightarrow E_0(t-\lambda t)-3Vr/2q$
where $r$ is the number of bands filled and $\lambda=E_0(t)V/3q$.
We call this the symmetric phase as all the symmetries of the system still remain intact in  this phase.
In this phase, the single  particle
energies just get scaled. But, these scalings are not uniform and depend  on
the bands filled.  However, the band gap never closes and 
the whole  fractal structure of the Hofstadter butterfly remains
intact.


The recursive relations for the  non-interacting case still remain valid for
the symmetric phase. The plot can be divided into $C$ and $D$ blocks.
Further, the $C$ block can be divided into sub-blocks and in  each sub-blocks,
the  whole energy spectrum is repeated in  a similar fashion as discussed in
the non-interacting case.

The scaling solution is not always the minimum energy solution. As 
discussed in section \ref{sec:II}, there are a large number of flux values
and filling fractions where there are phase transitions to  translational
symmetry broken states.

The plot  for flux per  plaquette  versus the maximum energy of  each band is
given in  Fig.~\ref{fig:7} for $V=1,2$ and $V=4$.  As seen from
Fig.~\ref{fig:7a} and Fig.~\ref{fig:7b}, for $V=1$, the Hofstadter butterfly,
like non-interacting case, can be divided into $C$ and $D$ blocks.  Viewing a
particular sub-block in $C$ block, as shown in Fig.~\ref{fig:7b}, we  see that
the form of the energy  spectrum  is not repeated inside this sub-block. Thus, the
plot is not self-similar.  For $V=4$, a larger portion of the fractal structure
gets destroyed compared to  $V=1$ as seen from Fig.~\ref{fig:7c}.

The fact that the fractal structure increasingly disintegrates with the 
increase in the interaction strength is a consequence of the increase in the number of
phase transitions with the increase in the interaction strength.  
The $C$ block contains energy bands near half filling which, as seen from
Fig.~\ref{fig:1}, have very high probability of phase transitions which
increases with $V$. These phase transitions, except at exact half filling,
break the translational symmetries of the system. Thus, the translational 
symmetry breaking phase transitions seem to play a crucial role in the 
destruction of the fractal structure. As we have stated earlier, this can
be expected on the basis of general arguments. 

For very dilute case, there are high  number of  phase transitions and they
affect the $D$ block. However, in this case, for small value of interaction
strength, the system is in symmetric phase for $p$ bands filled for flux $p/q$
as there  is a comparatively high  energy gap between  the  $p^{th}$ and
$(p+1)^{th}$ band. Thus, these high energy gaps in the $D$ block still remains
for small interaction strength and  will slowly vanish with increase in the interaction
strength.

\section{\label{sec:V}Landau fan diagram for the system in absence and presence of interactions}

\begin{figure*}[h!]
\centering
\includegraphics[scale=0.5,trim=1cm 1cm 0cm 2cm,clip]{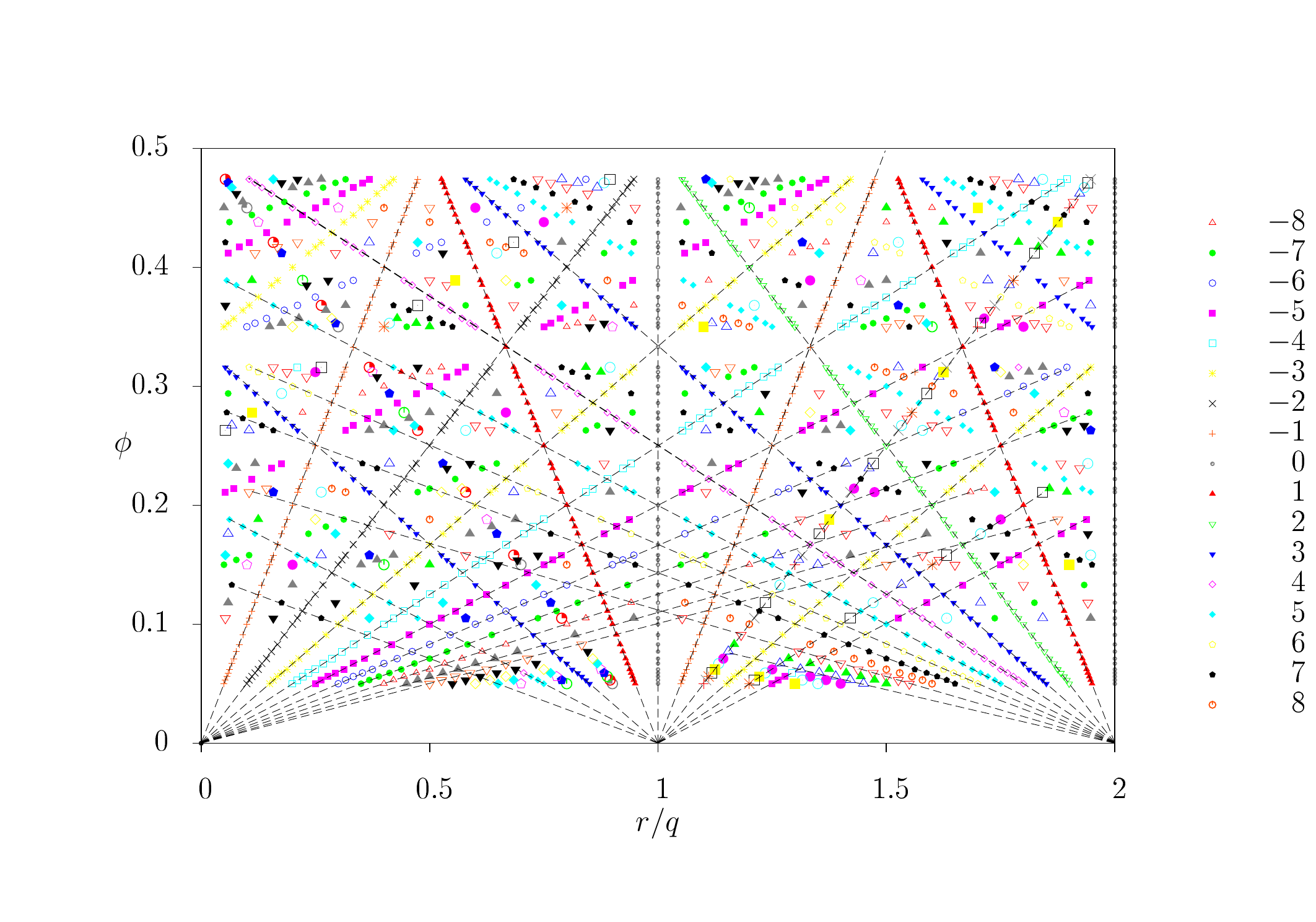}
\caption{\label{fig:8} (Color Online) Landau fan diagram for the non-interacting  case. In this figure
the  colorbar is  restricted to $t_r$ values from $-8$ to $8$ for convenience in plotting. This Landau fan
diagram is for $q\leq 20$.}
\end{figure*}

Experimental evidence for the Hofstadter butterfly has come from the
Landau fan diagram. 
Each gap in the Hofstadter butterfly can be characterized by two integer topological invariants $(t_r,~s_r)$
that satisfy the Diophantine equation \cite{donald1983,goldman2009}
  \begin{equation}
  r=t_rp+s_rq.
 \end{equation}
where $r$ labels the gap and the flux passing per plaquette is $\phi/\phi_0=p/q$, number  of particles per
unit cell is $r/q$. $t_re^2/h=-\sigma_H$ where
$\sigma_H$ is the Hall conductivity at the $r^{th}$ gap and $s_r$ is the change in the electron density when there is an
adiabatic change in the periodic potential \cite{donald1983}. 
The plot of the Hall conductivity with respect to the number of particles per unit cell and the magnetic flux passing
per plaquette is called the Landau fan diagram. Fig.~\ref{fig:8} shows the Landau fan diagram for 
the non-interacting case.

In Fig.~\ref{fig:8}, the points with the same Hall conductivities can be joined to give a straight
line which when extrapolated meets the x-axis at an integer point. This intercept 
gives the value of $s_r$ whereas the slope gives the value of $t_r$. 
In this figure, the  colorbar is  restricted to the values of $t_r$ from $-8$ to $8$ for convenience in 
plotting; the maximum  value  of $t_r$ for non-interacting case for $q\leq 20$ is $18$. Hence,
$q\leq 20$ is enough to show and analyze  the Landau fan diagram and realize that these
topological invariants indeed satisfy the Diophantine equation.

\begin{figure*}[!h]
\centering
\subfloat[]{\label{fig:9a}\includegraphics[scale=0.5,trim=1cm 1cm 0cm 2cm,clip]{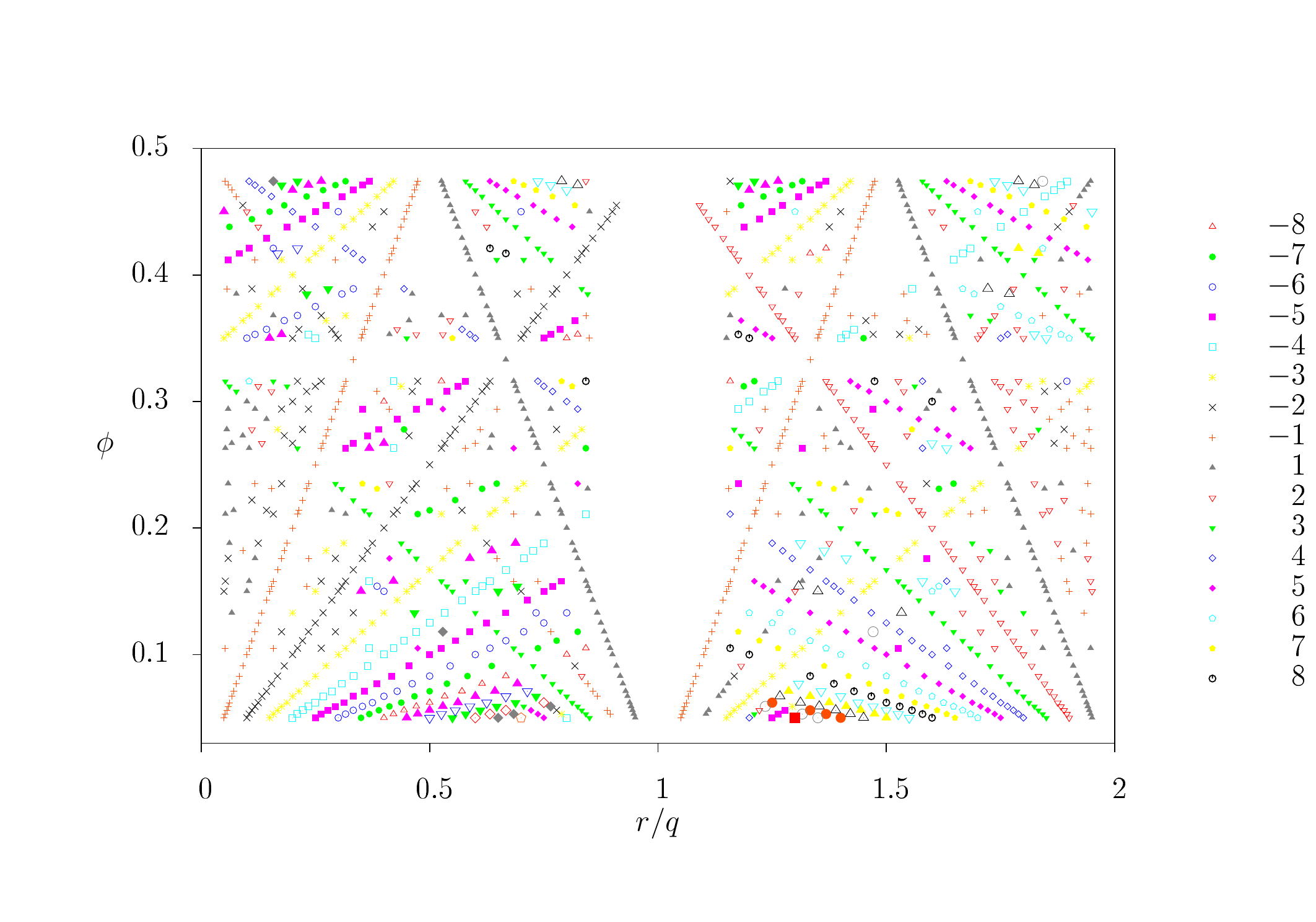}}\\
\subfloat[]{\label{fig:9c}\includegraphics[scale=0.5,trim=1cm 1cm 0cm 2cm,clip]{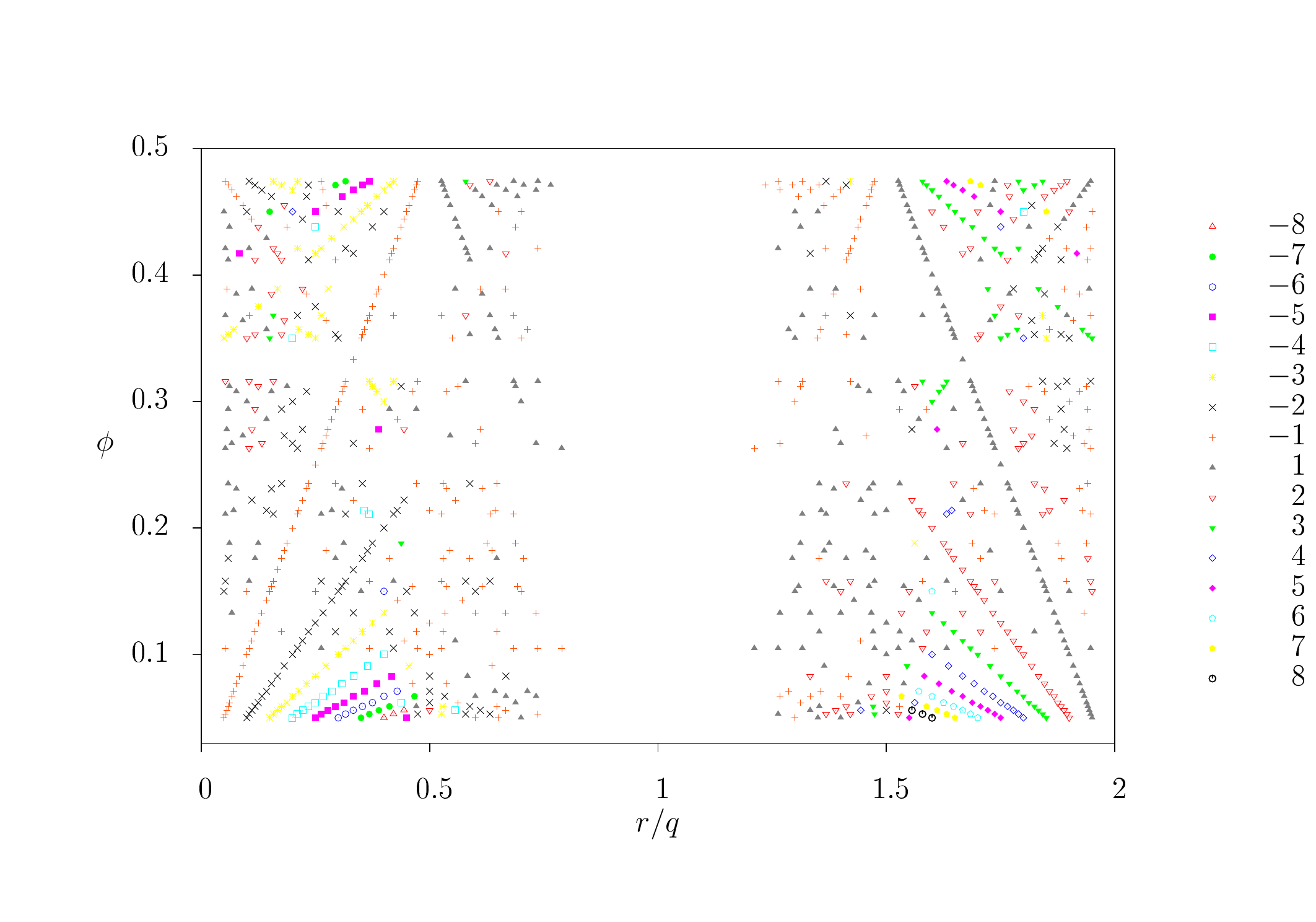}}
\caption{\label{fig:9}(Color Online) Landau fan diagram in the presence of interaction after removing the points where the  Hall
conductivity is zero for interaction strength (a) $V=1$ and (b) $V=4$. The  colorbar is  restricted to $t_r$ values from $-8$ to $8$ 
for convenience in plotting for (a).}
\end{figure*}

As mentioned earlier, many of the Landau transitions are accompanied by
topological transitions where the Hall conductivity of the system changes. In the presence  of interactions, the topological phase  transitions get  reflected in the Landau
fan diagram as shown in Fig.~\ref{fig:9} for $V=1,~2,~4$. 
Here the Landau fan diagram is plotted only for bands with non-trivial
topology, i.e. removing the points with zero  Hall conductivity.
In Fig.~\ref{fig:9a}, for $V=1$, though most of the points with the same Hall conductivities can be joined in
a straight lines but there are some points in these lines which have different Hall conductivities.
But in Fig.~\ref{fig:9c}, for $V=4$, the points with the same Hall conductivities cannot be  joined to form
a straight line as most of the points are scattered. This is due to the topological transitions accompanied
with the Landau phase transitions which  increases with the increase in the interaction strength.
Moreover, we can see that most of the region near half filling have a topological transition 
to zero Hall conductivity  as shown in Fig.~\ref{fig:9}.
This region also increases with the increase in the interaction  strength.
The maximum value of the Hall conductivity, considering all filled bands for all values of flux per plaquette 
of the form $p/q$ with $q\leq 20$, decreases with the increase of interaction strength. 
For example, in the absence of  interactions, the maximum value of the  Hall conductivity is $18e^2/h$, 
while for case of $V=1$ it is $15e^2/h$, for $V=2$ it is $11e^2/h$ and for $V=4$ the
maximum value of the Hall conductivity is $8e^2/h$.

Hence, from Fig.~\ref{fig:8} and Fig.~\ref{fig:9}, it is clear that due to the topological transitions accompanying
the phase transitions  to spatial symmetry breaking phases, the  points with same Hall conductivity in the Landau fan diagram
are more scattered and the number of transitions to  zero Hall conductivity 
near the half filling increases as
the interaction strength is increased.
In addition, on using the same  Diophantine equation as for the  non-interacting case, $s_r$ no longer remains
an  integer. 
For example, for $V=4$ and $r=2$, $t_2=0$, so $s_2=2/3$. Hence, the Diophantine
equation used for the non-interacting case is no more valid in the presence of interactions.

\section{\label{sec:VI}Conclusions}

We have studied spinless  fermions  on the honeycomb lattice with nearest
neighbor hopping and  nearest neighbor interaction in the presence of magnetic
field. The magnetic flux  per plaquette is of the form $p/q$  with $p,~q$ being
co-prime integers and $3\leq q\leq20$, $p< q$. We solve this interacting
problem in the mean field approximation for the filled band cases.  In
particular, we look for translation symmetry broken phases  and study the 
effect of these phase transitions on the Hofstadter butterfly and  the 
Landau fan diagram.

We find that a large number of the systems at different values of flux and
filling exhibit these transitions. Many of the transitions are also
topological, i.e. the symmetry breaking is accompanied by a change in the Hall
conductivity. When the number of transitions is plotted with respect to the
filling factor, we find that they are peaked near the dilute limit and near
half-filling. We have provided an explanation of this feature based on the
bunching of bands in the non-interacting system. The number of these phase
transitions increases with the increase in the interaction strength as expected.

The Hofstadter butterfly is generally understood as arising from the interplay
of the two length scales in the system, the periodicity of the potential and 
the magnetic length. The translation symmetry breaking introduces a third
length scale into the system and hence we expect a strong effect of it on 
the fractal structure. We have shown that this is indeed so. The self 
similarity structure of the energy spectrum disintegrates as a result of these
transitions. The amount of disintegration increases with the strength of the
interactions.

This result is with respect to the choice of the unit cell shown in 
Fig.\ref{fig:0}, corresponding to a particular pattern of translation symmetry
breaking. In this choice, the translational symmetry is broken only in the 
$\hat e_1$ direction. There could be states with different patterns of symmetry
breaking with lower energy and hence for many fillings and flux values, the 
translation symmetry could break at a smaller interaction strength than our
case. Thus if all patterns of translation symmetry breaking were taken into 
account, we may expect the Hofstadter butterfly to disintegrate at lower
interaction strengths than shown in this work.

Landau fan diagram is the experimental manifestation of the Hofstadter
butterfly.  We show that the change in the Hall conductivity in the transitions
gets reflected in the Landau fan diagram. The points with same Hall
conductivity no longer lie in  a straight line and are  rather scattered. On
increasing the interaction strength, there  are more number of topological
transitions and thus it  becomes  difficult to join the points with  same Hall
conductivities in  a straight  line as they get more scattered in the  Landau
fan diagram. 
In the presence of interaction, the Diophantine equation  used
for non-interacting case does not hold. 

Hence we have shown that interactions disintegrate the Hofstadter butterfly.
Further, the Landau fan diagram also drastically changes and the Diophantine
equation obeyed by the topological invariants in the non-interacting system
breaks down in the interacting system.

\end{document}